# A Dynamical Phase-Field Model for the Optical Properties of Ferroelectrics

Aiden Ross[1*], Anya Frazer[1], Venkatraman Gopalan[1], and Long-Qing Chen[1]

Ferroelectric materials are promising platforms for controllable photonic devices because of the strong coupling between their spontaneous polarization and optical properties. Yet, these materials remain challenging to design because of the close connection between the ferroelectric domain structure and optical response, which no existing theoretical approach can capture. Here, we develop a dynamical phase-field model that directly couples the ferroelectric order to the local optical response by introducing an electronic polarization field. This approach enables the prediction of the spatially resolved temperature- and wavelength-dependent optical properties in complex ferroelectric microstructures. Applying this method to $BaTiO_3$ thin films, we investigate the evolution of the local refractive index and electro-optic response under varying electric fields and temperatures. We show that the ferroelectric domain structure strongly modifies the local electro-optic response, exceeding 4000 pm/V near domain walls, several times larger than the bulk single crystal value ($r_{51}$=1300 pm/V). Our simulations quantitatively reproduce the electro-optic coefficient measured in $BaTiO_3$ on silicon films and capture the temperature-dependent behavior across multiple ferroelectric phase transitions, revealing the role of phase competition and coexistence in determining the electro-optic response. More broadly, this work establishes a general approach for predicting light-matter interactions in complex ferroelectric microstructures, enabling the computational design of ferroelectric materials for photonics.

[1]Department of Materials Science and Engineering and Materials Research Institute, The Pennsylvania State University, University Park, PA 16802

* Email Address: amr8057@psu.edu

## Introduction

Ferroelectrics are uniquely suited for controllable photonic devices, combining a non-centrosymmetric crystal structure with a switchable electrical polarization. The non-centrosymmetric structure can support second-order optical nonlinearities[1], including the linear electro-optic effect, second-harmonic generation, sum- and difference-frequency generation, enabling applications in electro-optic modulators[2], frequency conversion, and quantum photonics[3–6]. Moreover, their piezoelectric properties enable large acousto-optic properties for acousto-optic modulators[7,8] and beam steering[9]. This switchable polarization allows for the precise, non-volatile control of local optical properties using applied electric fields, which serves as the basis for non-volatile phase-shifters[10], reconfigurable waveguides[11], and quasi-phase-matching[12,13], making ferroelectrics a versatile class of materials for photonics.

Despite these promising properties, ferroelectrics exhibit substantial complexity arising from the strong couplings between the ferroelectric polarization and the electrical and mechanical degrees of freedom, which often lead to the formation of complex mesoscale domain structures that strongly influence their optical response. For example, in $BaTiO_3$, nanoscale ferroelectric domains lead to nanoscale variations in the local optical properties[14] that are history-dependent[10,15]. As a result, the effective optical properties of ferroelectric thin films can substantially differ from ideal single crystals. Although $BaTiO_3$ possesses one of the highest electro-optic coefficients as a bulk single crystal ($r_{51} = 1300\ \mathrm{pm/V}$)[16], when grown as a thin film the effective electro-optic coefficients are often smaller than the bulk value[17,18], and show a significant variability (ranging from $r_{eff} = 200\,\frac{\mathrm{pm}}{\mathrm{V}} - 1080\,\frac{\mathrm{pm}}{\mathrm{V}}$). Therefore, a major challenge remains to harness this complex behavior to design ferroelectric materials for photonics.

While first-principles calculations[19–22] and phenomenological models[23,24] are critical to provide a fundamental understanding of the electro-optic properties, they are inherently limited in their ability to capture the mesoscale dynamics that often govern the optical properties[22]. The phase-field method is naturally suited to predicting the formation and dynamics of mesoscale domain structures,[25] however, existing formulations lack the electronic degrees of freedom responsible for near-infrared to visible optical properties. In our previous work, we developed a thermodynamic theory of optical properties in ferroelectrics, which derives the optical properties by separating the polarization into lattice and electronic contributions and describing their energetic couplings[26]. The lattice polarization arises from the ionic displacements that govern the ferroelectric order, while the electronic polarization describes the field-induced distortions of the electronic orbitals. Because the lattice polarization is associated with ionic motion, its effective mass is too large to respond at optical frequencies, thus the optical properties arise from the dynamics of the electronic polarization. Although prior phase-field studies have attempted to compute electro-optic properties, these only considered the lattice contributions, which fail to accurately capture the optical response[27]. For example, following this approach would predict a room-temperature refractive index of 54 for $BaTiO_3$, whereas the experimental value is approximately 2.3[28].

Here, we develop a dynamical phase-field model for the optical properties of ferroelectric materials. The model describes the coupled evolution of the lattice polarization, electronic polarization, elastic fields, and electromagnetic fields, enabling the direct simulation of ferroelectric domain evolution and local optical properties across experimentally relevant length scales. From this general formulation, we derive expressions for the local refractive index and electro-optic coefficients under periodic optical electric fields, allowing spatially resolved optical properties to be computed directly from the evolving microstructure. As a case study, we apply this approach to $BaTiO_3$ thin films, at and near 0% strain, representative of relaxed films integrated onto silicon, and investigate the evolution of the optical properties as a function of electric-field orientation and temperature. We demonstrate a strong quantiative agreement with experimental measurements and reveal new microstructural contributions to the electro-optic response. These results establish a new theoretical approach for understanding and designing ferroelectric materials for photonic applications.

## Phase-Field Model Formulation

### Timescales in Ferroelectric Materials

Several mechanisms contribute to the dielectric susceptibility of ferroelectrics, each with a characteristic timescale, as illustrated in Figure 1. The total dielectric response is the sum of all these different contributions that remain active at a given timescale. Ignoring free-carrier mediated effects, at low frequencies, all three mechanisms, electronic polarization dynamics[1], lattice polarization dynamics[29], and collective ferroelectric domain dynamics[30], contribute to the dielectric susceptibility. In the near-infrared to visible spectrum, however, collective ferroelectric domain dynamics and lattice polarization dynamics are far past their resonance / relaxation frequency and become effectively frozen. As a result, the optical response is almost entirely governed by electronic polarization dynamics.

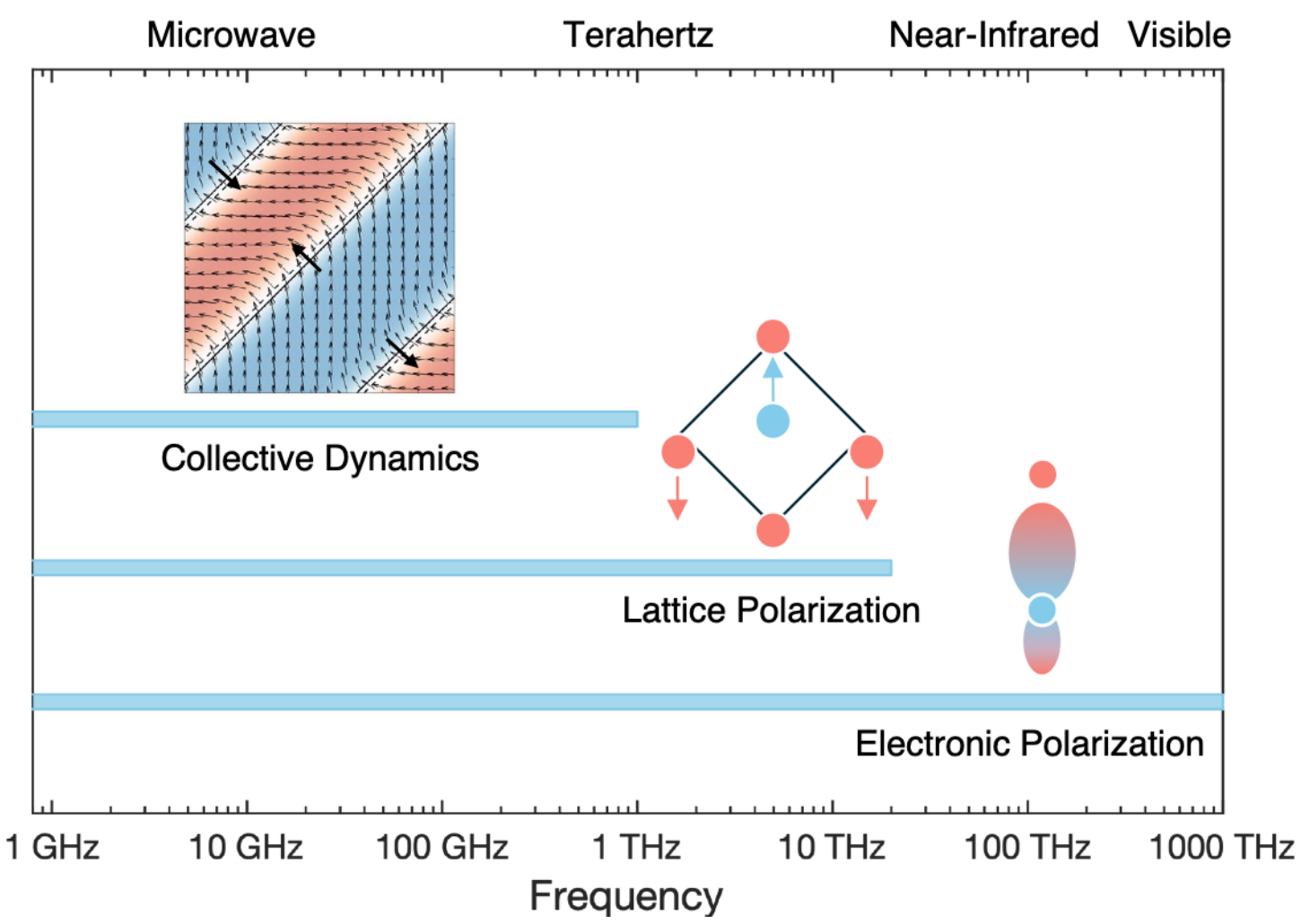

**Figure 1. Characteristic frequency scales in the dynamics of ferroelectric materials.** Collective domain dynamics, lattice polarization dynamics, and electronic polarization dynamics all contribute to the dielectric response, but at optical frequencies the only the electronic polarization can respond.

Changes in the optical properties therefore involve changes to the electronic polarization energy landscape. These changes can occur directly through anharmonicities in the electronic energy landscape, which are responsible for optical second-harmonic generation and the electronic electro-optic effect[1,26]. These changes can also occur indirectly through coupling to another field. For example, although the lattice polarization cannot directly follow an optical electric field, it can indirectly modify the optical response through its coupling to the electronic polarization[23]. Similarly, strain and stress can modify the electronic energy landscape through electromechanical coupling[24].

The electro-optic effect provides an important example of this interaction. In this case, a low-frequency electric field modifies the lattice polarization, which, through lattice-electronic coupling, perturbs the electronic polarization energy landscape and alters its interaction with an optical electric field[26]. An additional contribution may also arise from domain dynamics. Here, a low-frequency electric field can shift the position of ferroelectric domain walls, modifying the spatial distribution of local polarization and stress. Because electronic polarization is coupled to each of these quantities, these dynamics can generate an emergent contribution to the electro-optic response associated with the evolving domain structure. A complete description of the optical response of ferroelectric materials must account for both the electronic polarization and the influence of slower-evolving fields that determine the local electronic energy landscape.

**General Model**

In the phase-field method, the thermodynamic state of the ferroelectric system is described by a free energy functional that depends on the lattice polarization field, $\boldsymbol{P^L}(\boldsymbol{x},t)$ , the electronic polarization field, $\boldsymbol{P^e}(\boldsymbol{x},t)$, stress field, $\boldsymbol{\sigma}(\boldsymbol{x},t)$), and electric field $\boldsymbol{E}(\boldsymbol{x},t)$. The total free energy functional may be written as

$$F\left(T,P_i^L,P_i^e,\sigma_{ij},E_i\right) = \int\left[f^L(T,P_i^L)+f^e(T,P_i^e)+f^{Le}(P_i^L,P_i^e)+f^{grad}\left(\nabla_i P_j^L\right)+f^{elas}\left(P_i^L,P_i^e,\sigma_{ij}\right) + f^{elec}(P_i^L,P_i^e,E_i)\right]dV \tag{1}$$

where $f^L$ describes the intrinsic free energy of the lattice polarization[31], and $f^e$ describes the free energy of the electronic polarization when $\sigma_{ij}=0$ and $P_i^L=0$. $f^{Le}$describes the energetic coupling between the lattice and electronic polarization[26]. The gradient energy contribution is denoted by $f^{grad}$ where we assume that the gradient energy of the electronic polarization is negligible, $f^{elas}$ is the elastic energy, and $f^{elec}$is the electrostatic energy.

The evolution of the lattice and electronic polarization follows the polarization dynamics equation[32,33]

$$\mu_{ij}^L\frac{\partial^2 P_j^L}{\partial t^2}+\gamma_{ij}^L\frac{\partial P_j^L}{\partial t}=-\frac{\delta F}{\delta P_i^L} \tag{2}$$

$$\mu_{ij}^e\frac{\partial^2 P_j^e}{\partial t^2}+\gamma_{ij}^e\frac{\partial P_j^e}{\partial t}=-\frac{\delta F}{\delta P_i^e} \tag{3}$$

The evolution of the elastic displacement is governed by the elasto-dynamic equation[33]

$$\rho\frac{\partial^2 u_i}{\partial t^2}=\nabla\cdot\left(\sigma_{ij}+\beta\frac{\partial\sigma_{ij}}{\partial t}\right) \tag{4}$$

and the evolution of the electric field and magnetic field are governed by Maxwell's equations

$$\nabla\times H_i=\frac{\partial D_i}{\partial t}+J_i \tag{5}$$

$$\nabla\times E_i=-\frac{\partial B_i}{\partial t} \tag{6}$$

$$\nabla_i\cdot B_i=0 \tag{7}$$

$$\nabla_i\cdot D_i=\rho_f \tag{8}$$

Equations 1-8 provide a fully coupled description of the lattice, electronic, elastic, and electromagnetic degrees of freedom involving coupled light-matter interactions in ferroelectric materials. A schematic diagram of the governing equations is shown in Figure 2.

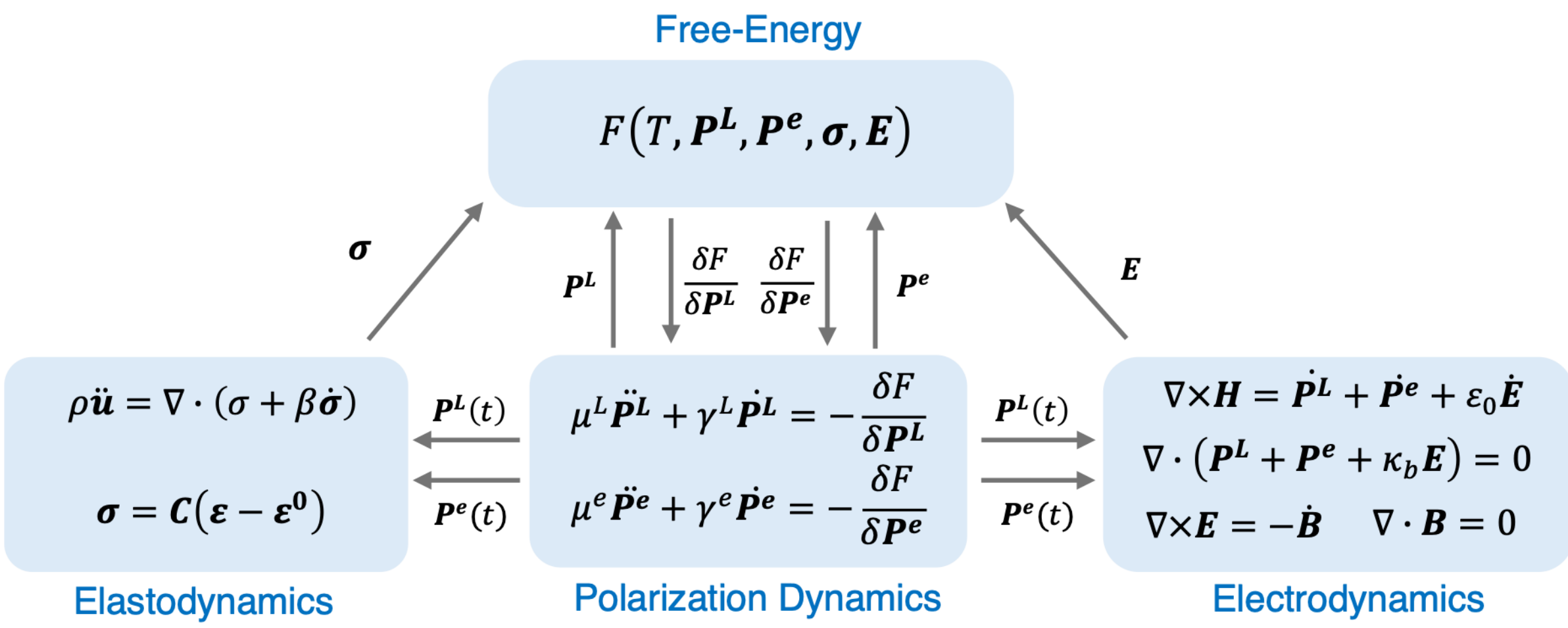

**Figure 2. Diagram of the coupled dynamics within the phase-field model.** The free-energy functional describes the state of the system using the temperature, lattice polarization, electronic polarization, stress, and electromagnetic fields. Functional derivatives of the free energy provide the driving forces for the lattice and electronic polarization dynamics. The evolving polarization fields act as source terms in the electrodynamic equations, and the polarization is coupled to the evolving elastodynamics through the eigenstrain. The resulting polarization, stress, and electric fields feed back into the free-energy functional, creating a spatiotemporal description of the ferroelectric domain evolution and optical properties.

### Probing Local Properties via Perturbation Theory

The large separation between the timescales of the lattice polarization dynamics and electronic polarization dynamics (Figure 1) allows the fully coupled equations to be simplified. Here, the evolution of the lattice polarization may be treated separately, with the electronic polarization assumed to instantaneously equilibrate in response to changes in the lattice polarization and stress fields. This allows the local optical properties to be captured without explicitly resolving the femtosecond dynamics of the electronic polarization, making the approach far more computationally feasible.

Under these conditions, we may adopt the relaxational approximation for the lattice polarization and assume that the mechanical displacement, electrostatic potential, and electronic polarization instantaneously reach equilibrium in response to changes in the lattice polarization. The evolution equations simplify to

$$\gamma_{ij}^L \frac{\partial P_j^L}{\partial t} = -\frac{\delta F}{\delta P_i^L} \tag{9}$$

$$\nabla_i \sigma_{ij} = 0 \tag{10}$$

$$\nabla_i \left( \epsilon_0 \kappa_{ij}^b E_j + P_i^L \right) = 0 \tag{11}$$

Where $\kappa_{ij}^b$ is the background dielectric constant tensor which contains contributions from the electronic dielectric susceptibility and the vacuum permittivity.

Solving equations 9-11 yields the local lattice polarization and stress fields throughout the ferroelectric microstructure which determine the local free-energy landscape of the electronic polarization and therefore the local optical properties. The near-infrared and visible optical response can be calculated by solving the electronic polarization dynamics (equation 3) using a perturbation expansion[1]

$$\mu_{ij}^{e}\frac{\partial^{2}P_{j}^{e}}{\partial t^{2}}+\gamma_{ij}^{e}\frac{\partial P_{j}^{e}}{\partial t}+\frac{1}{\epsilon_{0}}B_{ij}^{e}P_{j}^{e}+\cdots=E_{i}. \tag{12}$$

Where $B_{ij}^{e}(x_i)=\epsilon_0\left(\frac{\partial^2 f(x_i)}{\partial P_i^e \partial P_j^e}\right)$ is the electronic dielectric stiffness which is related to the local curvature of the electronic polarization energy surface. Assuming a monochromatic electric field $E_i(t)=E_i(\omega)e^{-i\omega t}$, and a polarization $P_i^e(t)=P_i^e(\omega)e^{-i\omega t}$ equation 12 simplifies to,

$$P_{i}^{e}(\omega)=\left[-\omega^{2}\mu_{ij}^{e}-i\omega\gamma_{ij}^{e}+\frac{1}{\epsilon_{0}}B_{ij}^{e}\right]^{-1}E_{j}(\omega), \tag{13}$$

and the linear optical dielectric susceptibility is given by

$$\chi_{ij}^{e,(1)}(x_i,\omega)=\tilde{n}_{ij}^{2}(x_i,\,\omega)-\mathbb{I}=\left[B_{ij}^{e}(x_i)-\epsilon_0\left(i\omega\gamma_{ij}^{e}+\omega^{2}\mu_{ij}^{e}\right)\right]^{-1} \tag{14}$$

where $\mathbb{I}$ is the identity tensor. The optical dielectric susceptibility is generally frequency dependent and complex, with the real part determining the phase velocity of light and the imaginary part describing optical absorption. The damping tensor $\gamma_{ij}^{e}$ introduces the imaginary component to the susceptibility and governs the absorption in the material. The effective mass tensor $\mu_{ij}^{e}$ and electronic dielectric stiffness determines the resonance frequency $\omega_0^2=\frac{B^e}{\mu\epsilon_0{}^e}$. At frequencies much smaller than the resonance frequency the electronic dielectric constant becomes approximately frequency independent and is related to the inverse of the electronic dielectric stiffness, $\chi_{ij}^{e}=\left(B_{ij}^{e}\right)^{-1}$. Because $B_{ij}^{e}$ is determined by the local lattice polarization and stress through the free energy, the optical dielectric susceptibility directly inherits the spatial variations of the ferroelectric microstructure.

Because the electronic polarization is assumed to instantaneously reach equilibrium, relative to the typical timescales of the lattice polarization, the local optical properties can be evaluated at every step of the ferroelectric domain evolution. This allows for the prediction of the coupled evolution of the ferroelectric microstructure and the associated refractive index distributions under applied electric fields. Using this approach, the local electro-optic coefficients are numerically evaluated by applying a small external electric field to the system and evaluating the difference of the inverse dielectric tensor between these two conditions,

$$r_{ijk}(\omega,x_i)=\left(\frac{\partial B_{ij}(\omega,x_i)}{\partial E_k}\right)\cong\frac{B_{ij}^{[2]}(\omega,x_i)-B_{ij}^{[1]}(\omega,x_i)}{E_k^{[2]}-E_k^{[1]}}, \tag{15}$$

where $B_{ij}$ is the inverse optical dielectric tensor given by, $B_{ij}=\left(\frac{1}{n^2(\omega)}\right)_{ij}=\left[\chi^{e,(1)}(\omega)+\mathbb{I}\right]_{ij}^{-1}$, and $E_i$ is the uniform electric field applied to the simulation. For simplicity we neglect the imaginary component of the electronic dielectric tensor. Although this current work is focused on the linear electro-optic response, this same approach can be extended to compute higher order electro-optic coefficients, nonlinear optical susceptibilities, thermo-optic coefficients, piezo-optic effects, and other optical responses that emerge from the coupled evolution of the ferroelectric microstructure and electronic polarization.

## Results

To demonstrate the capabilities of the proposed theoretical approach, we investigate the electro-optic response of $BaTiO_3$ thin films, which are promising materials for integrated silicon photonics and quantum photonic technologies because of their large electro-optic coefficients and compatibility with silicon[34–36]. We examine how the ferroelectric domain structure determines the refractive index, then track the evolution refractive index under an applied electric field, leading to the electro-optic effect. We further examine how this electro-optic response changes with the orientation of the applied electric field and finally track the evolution of the electro-optic response with temperature demonstrating a quantitative agreement with experimental measurements.

### Local Refractive Indices

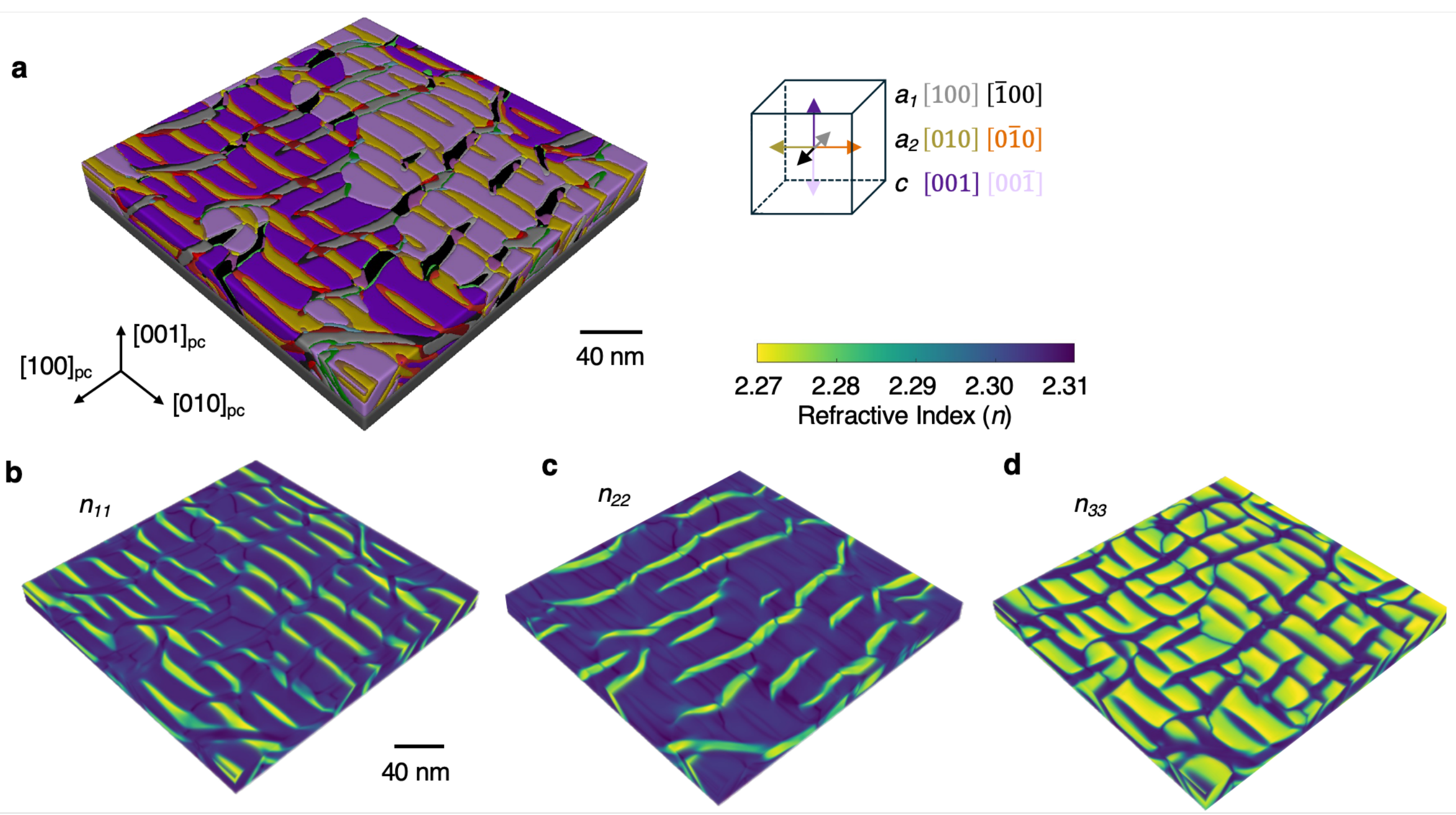

**Figure 3.** (a) Ferroelectric domain structure of $BaTiO_3$ at 350K, under a 0% biaxial strain, showing variants $a_1$, $a_2$, and c along $[100]_{pc}$, $[010]_{pc}$, and $[001]_{pc}$, respectively. (b-d) corresponding local refractive indices $n_{11}$, $n_{22}$, and $n_{33}$ at $\lambda = 1550$ nm.

Figure 3a shows the ferroelectric domain structure of $BaTiO_3$ at 0% strain and 350K. The domain structure consists of three tetragonal domain variants, $a_1$, $a_2$, and c, with spontaneous polarizations aligned parallel to the [100], [010] and [001] directions respectively. The corresponding local refractive indices, $n_{11}$, $n_{22}$, and $n_{33}$ are presented in Figures 3b-d. Since the ferroelectric polarization defines the optic axis for the refractive index, the orientation of the optical indicatrix rotates with the ferroelectric polarization. Since $BaTiO_3$ is a negative uniaxial crystal, the optic axis exhibits a smaller refractive index than the other two directions. Consequently, the $a_1$ domains possess a smaller $n_{11}$ refractive index, the $a_2$ domains possess a smaller $n_{22}$ refractive index and the c domains possess a smaller $n_{33}$ refractive index. As a result, the ferroelectric domain structure leads to abrupt changes in the refractive index across the domain walls.

## Evolution of optical properties under applied electric fields

The relationship between the ferroelectric domain structure and the refractive index provides the basis for understanding the electro-optic response. As the ferroelectric domain structure evolves under an applied electric field, so does the local refractive index. Therefore, the total electro-optic response reflects both the change in the refractive index within each individual domain and the change in the refractive index caused by domain wall motion or the nucleation of new phases.

Figure 4 shows the coupled evolution of the ferroelectric polarization, refractive index, and electro-optic response of $BaTiO_3$ under an applied electric field along the $[100]_{pc}$ direction. Figure 4a shows the polarization-electric field hysteresis loop, and Figure 4b presents the refractive index-electric field hysteresis loop. Figure 4d-f plots the evolution of the ferroelectric polarization distribution (top), the refractive index distribution (middle), and the electro-optic coefficient distribution (bottom) under an applied electric field.

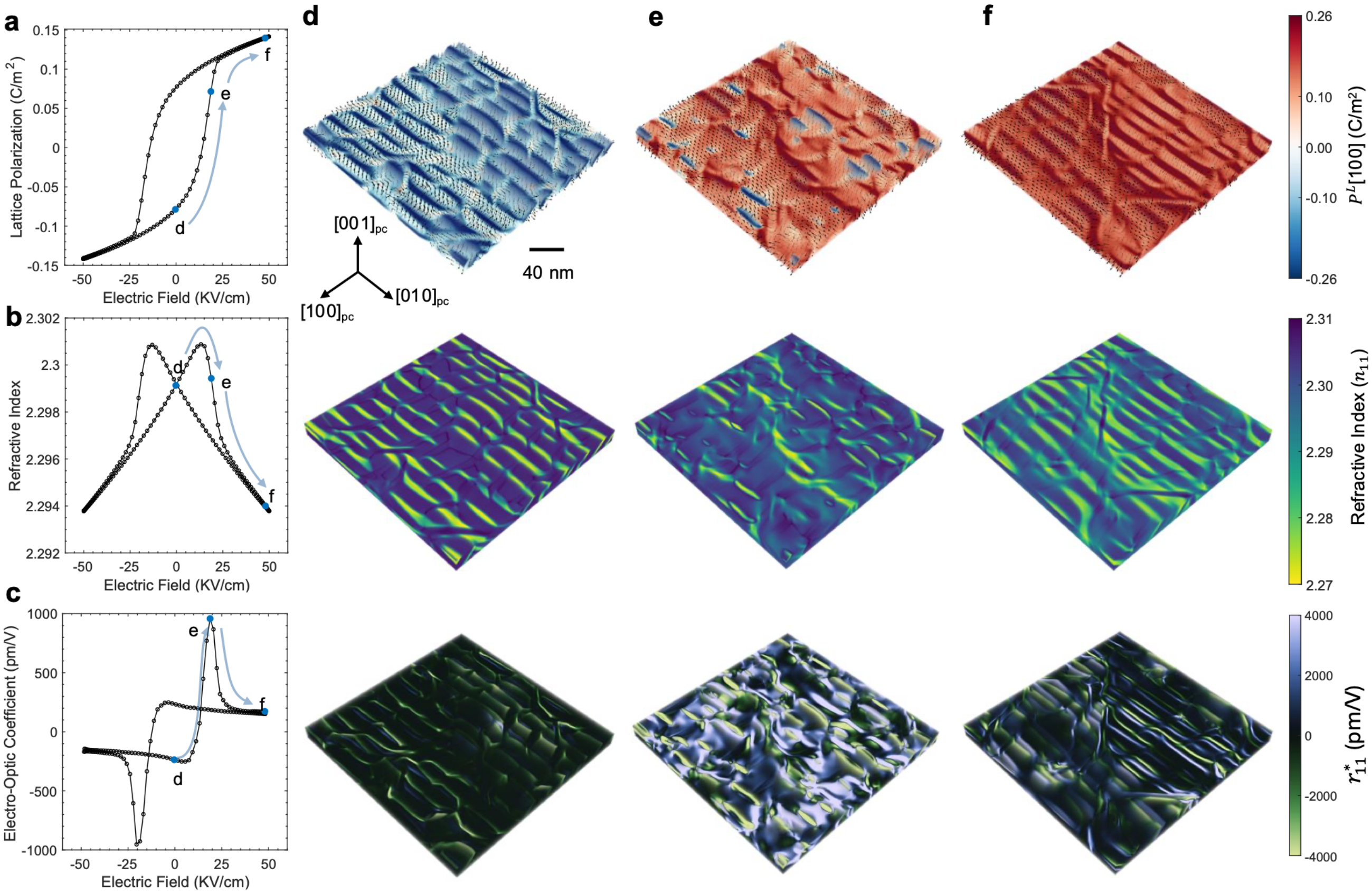

**Figure 4. Evolution of ferroelectric domain structure and local refractive indices under an applied electric field.** a) polarization electric field hysteresis loop, b) refractive index electric field hysteresis loop, c) electro-optic coefficient electric field hysteresis loop, d-f) local electrical polarization, refractive index, and electro-optic coefficient distribution under different applied electric fields.

The initial state (Figure 4d) consists of a mixed $+c/-a_1$ domain structure. Upon applying an electric field along the $[100]_{pc}$ direction, the $-a_1$ domains shrink while the $+c$ domains expand, leading to an increase in $n_{11}$. At approximately 20 kV/cm (Figure 4e), the polarization begins to switch, causing $+a_1$ to nucleate near the bottom of the film while the $-a_1$ domains disappear. After switching, $+a_1$ domains have a larger volume fraction compared to the $+c$ leading to an abrupt change to smaller $n_{11}$ values. With a further increase in electric field (Figure 4f), the $+a_1$ domains grow, causing $n_{11}$ to decrease. Finally, when the electric field is removed the polarization has switched, while the refractive index relaxes close to its initial value producing the refractive-index hysteresis loop shown in Figure 4b.

From these field-dependent refractive index changes, we can directly quantify the local electro-optic effect, which reaches values greater than 4000 pm/V near the ferroelectric domain walls, although the film on average possesses an electro-optic coefficient around 230 pm/V. Moreover, during switching, these domain wall contributions grow, causing the average electro-optic effect to approach 1000 pm/V, but only over a narrow electric field window (Figure 4c and Figure 4e).

## Influence of electric field orientation

We further investigate the electro-optic effect of $BaTiO_3$ thin films under different applied electric field orientations as shown in Figure 5. Figures 5a-c show the polarization, refractive index, and electro-optic effect hysteresis loops for electric field orientations of 0°, 22.5° and 45°, where the angle is measured between the applied electric field and the $[100]_{pc}$ direction in the plane of the film. For each orientation, the plotted quantities correspond to the in-plane tensor components along the applied electric field direction, representing the effective refractive index and electro-optic response experienced by a $TE_0$ mode whose optical polarization is parallel to the modulating field.

The corresponding domain structures, refractive index and electro-optic coefficients as $E \rightarrow 0$ kV/cm are shown in Figures 5d-f for the 0°, 22.5° and 45° electric field orientation. In the tetragonal phase of $BaTiO_3$ the polarization prefers 6 possible orientations along the $<100>_{pc}$ family of directions and the relative volume fraction of these domains is controlled by the poling direction which favors domains that align with the external electric field direction. Although the tetragonal domains are favored, domains that differ from these preferred orientations may also be present, due to electrical and mechanical compatibility conditions between the domains. The corresponding refractive index distribution is shown in the middle (Figure 5d-f), with the average refractive index of the film decreasing as the poling direction grows closer to 45°. In $BaTiO_3$ the refractive index is the smallest along the spontaneous polarization direction. As the poling direction rotates towards the 45° direction, the light increasingly samples domain variants whose polarization direction aligns with the optical electric field, thus reducing the effective refractive index. In addition, the 45° orientation stabilizes some small orthorhombic regions with even lower refractive indices.

Under these conditions, the electro-optic response is inhomogeneous, showing large enhancements near the ferroelectric domain walls similar to Figure 4 and by changing the applied electric field direction these contributions can be further activated. For example, under the 0° orientation electric field, these enhancements are present primarily between the $a_1$ and c domains with the $a_2$ domain walls remaining relatively inactive. This is due to two reasons: first the 0° electric field will not produce a substantial driving force for the $a_2$ domains to grow at the expense of the c domains and second, the $a_2$ and c domains will possess roughly equal refractive indices which limits the effective domain wall contribution. As the applied electric-field angle increases near 45° the domain walls from both the $a_1$/c and $a_2$/c domains become active, increasing the domain wall contribution to the electro-optic response. Beyond the domain wall contributions, electro-optic response in the bulk of each domain will also change with the applied electric field direction. In general, perovskite ferroelectrics tend to have a higher dielectric susceptibility when the electric field is perpendicular to the electrical polarization than when it is parallel which can be thought of as having a higher susceptibility to polarization rotation. As the electric field is rotated to the 45° direction, the electric field can induce polarization rotation in both the $a_1$ and $a_2$ domains leading to a larger domain contribution.

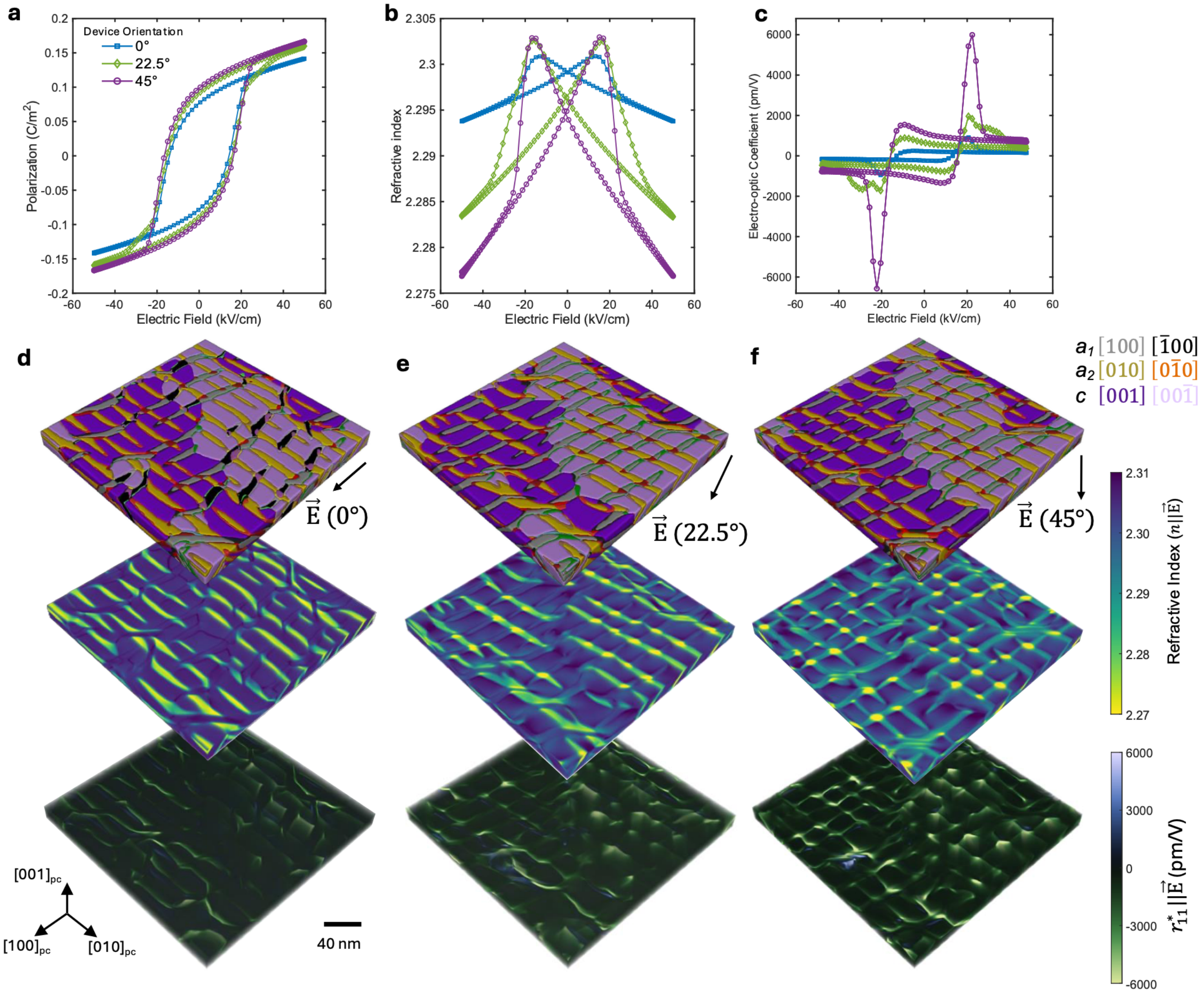

**Figure 5. Electro-optic response under different electric field orientations.** (a-c) Hysteresis loops of polarization, refractive index, and electro-optic coefficient respectively, with each quantity projected along the direction of the applied electric field. (d-f) Domain structure (upper) local refractive index distribution (middle) and electro-optic coefficient distribution (lower) 0°, 22.5°, and 45° applied electric fields. In the visualizations, the plotted quantities are aligned with the applied electric fields.

Beyond the local enhancement of the electro-optic effect near the domain walls, the film-averaged electro-optic response shows a large transient increase during the switching process which is also dependent upon the applied electric field angle. Near the coercive field, the effective electro-optic coefficient reaches approximately 1000 pm/V for the 0° orientation, 2000 pm/V for the 22.5° orientation, and 6000 pm/V for the 45° orientation, where the latter exceeds the largest bulk single crystal coefficient of $BaTiO_3$ by nearly a factor of 5 (Figure 5c). This trend further reflects the growing fraction of the microstructure that participates in switching, where under the 0° orientation, the field primarily reconfigures the $a_1$ / $c$ domain population, whereas at 45°, the $a_1, a_2,$ and $c$ domains are all active, so a large volume fraction of the film can simultaneously participate in the switching dynamics.

## Influence of temperature

Next, we examine how temperature influences the electro-optic response in $BaTiO_3$ under different poling directions. As a bulk single crystal, $BaTiO_3$ undergoes three phase transformations from cubic to tetragonal at 400K, tetragonal to orthorhombic at 270K, and from orthorhombic to rhombohedral at 200K. In thin films, however, the mechanical boundary conditions modify the stability of the ferroelectric phase, shifting the phase transition temperatures and stabilizing lower symmetry monoclinic phases that are absent in the bulk phase diagram. To compare with experiments on Si-integrated films, these simulations include small residual tensile strain of 0.15% characteristic of $BaTiO_3$/Si (see Methods).

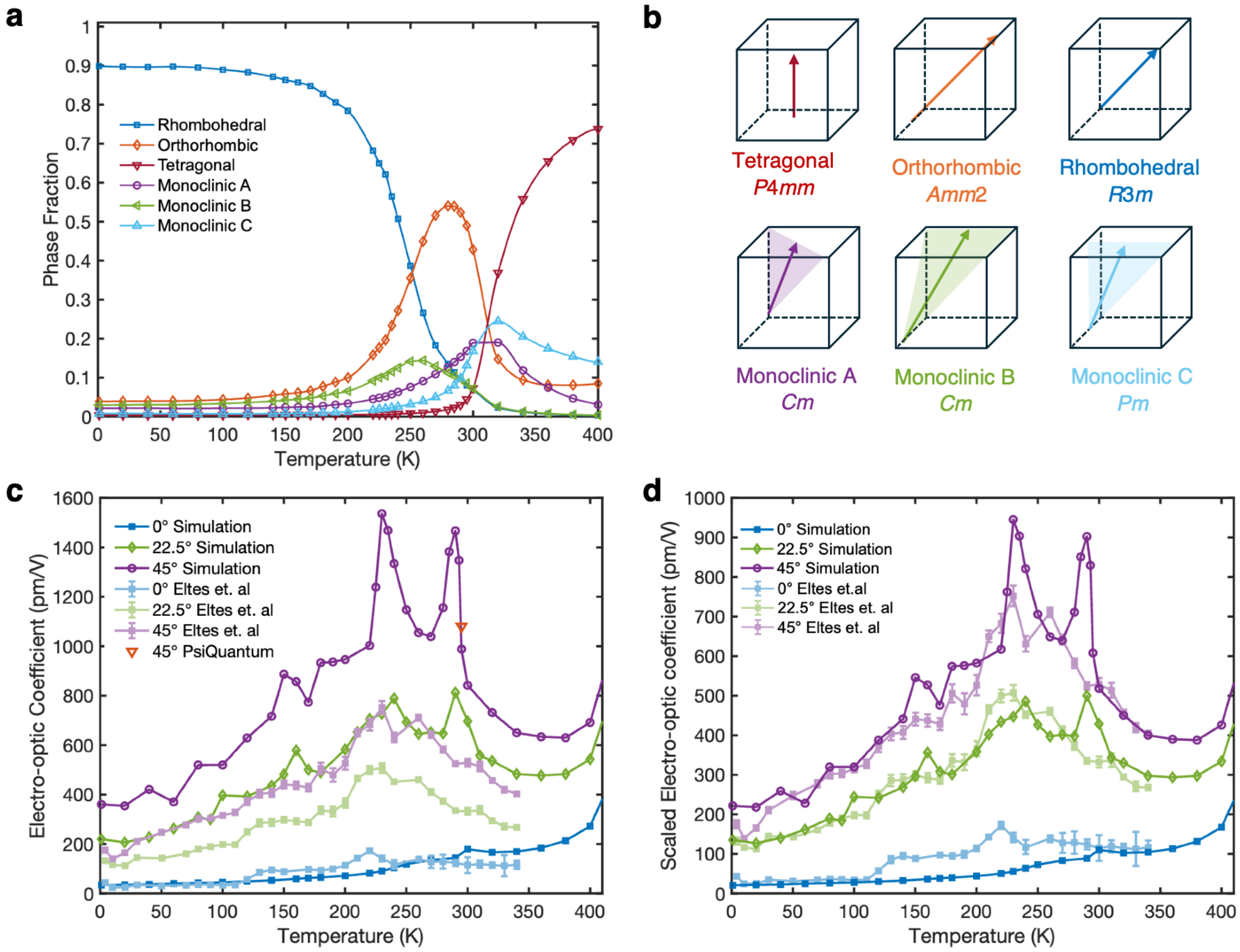

**Figure 6. Temperature Dependent Electro-Optic Response** (a) The phase-fractions as a function of temperature in $BaTiO_3$ under 0.15% strain (b) Schematic diagrams of the ferroelectric phases in (a), with arrows indicating the polarization direction and shaded planes indicating the plane within which the polarization is confined for the monoclinic phases. (c) The simulated temperature dependent effective electro-optic coefficient along different crystal orientations from 1 K – 400 K (darker) compared to experimental measurements from Eltes et al.[18] (lighter) and the room temperature measurement from PsiQuantum[36] (orange triangle) (d) The temperature dependent electro-optic

response with the simulated data scaled by a constant factor chosen to match the 0° measurement of Eltes et al. at 300K.

The phase fractions extracted from the simulations are shown in Figure 6a. Although the simulations generally follow the bulk sequence of ferroelectric phase transitions, the transitions become more diffuse, with broader regions of phase coexistence and the emergence of monoclinic bridging phases. Near the tetragonal to orthorhombic transition, the monoclinic c phase ($M_c$) becomes prominent, whereas the monoclinic b phase ($M_b$) is stabilized near the orthorhombic to rhombohedral phase transition. Schematic diagrams of the polarization direction for each monoclinic phase are shown in Figure 6b. The relative fractions of these phases are history dependent and vary with poling conditions (Supplementary Note 2).

Figure 6c presents the calculated electro-optic response as a function of temperature for different applied electric field orientations, compared with experimental measurements from Eltes et al.[18] and from Alexander et al. (PsiQuantum)[36]. The simulations show a quantitative agreement with the measurements from PsiQuantum (1080 pm/V at 295K vs 988 pm/V at 295 K) suggesting that their films are approaching the intrinsic response of the $BaTiO_3$. The measurements from Eltes et al. are systematically lower in magnitude, however the temperature dependence and the relative magnitudes among the three electric field orientations match well with the simulation, as shown by scaling the simulated response by a constant factor to match the 0° measurement at 300 K (Figure 6d). This discrepancy may be explained by the presence of interfacial dead layers or incomplete poling, that may reduce the magnitude of the measured response without altering the sequence of ferroelectric phase transitions that describe the temperature dependence.

By comparing the phase fractions and electro-optic response as a function of temperature we find that the largest electro-optic coefficients occur within the vicinity of the tetragonal to orthorhombic and orthorhombic to rhombohedral phase boundaries. In these regions, the electro-optic coefficient reaches values approaching 1500 pm/V, approximately 2.5 times larger than those obtained within the center of the tetragonal phase. The enhancement of the electro-optic response near the phase boundaries suggests that the electro-optic design principles should not focus on stabilizing one phase or crystal structure but instead should prioritize engineering the conditions where multiple ferroelectric phases are in close competition. Another potential strategy is to create kinetic pathways which allow a low energy conversion between ferroelectric states with different refractive indices. Under these conditions, the presence of multiple ferroelectric phases and the emergence of low-symmetry monoclinic phases increase the susceptibility of the lattice polarization to applied electric fields thereby enhancing the electro-optic response. We also note that the electrical poling process modifies the phase fraction of the material and shifts the observed ferroelectric phase transition temperatures (see Supplementary Note 2).

## Discussion

In this work, we developed a dynamical phase-field model that directly predicts the local optical response in ferroelectric microstructures. By introducing an electronic polarization field, the model extends the current phase-field descriptions of ferroelectrics to describe optical properties, enabling the calculation of spatially resolved refractive indices and electro-optic coefficients. By applying this framework to $BaTiO_3$ thin films, the simulations reproduce the experimentally observed temperature- and electric-field-angle dependence of the electro-optic response and identify the nanoscale structural motifs that enhance the electro-optic effect.

Existing theoretical approaches based on density functional theory accurately predict the intrinsic optical tensors of homogeneous crystals, but they are generally limited to length scales much smaller than those of typical ferroelectric domain structures[19]. On the other hand, other optical models, which typically construct multidomain systems using the bulk electro-optic tensor and assign it to artificially generated domains, can provide useful insights into the anisotropic electro-optic response, but these models assume that each domain behaves as an isolated crystal and ignore potential interactions between domains[37,38]. By consequence, the electro-optic tensors measured from bulk single crystals are often insufficient to describe the local optical response of ferroelectric thin films.

Our proposed framework bridges this gap by capturing both the evolution of the ferroelectric microstructure and the corresponding evolution of the local optical response, thereby tracking the optical response through domain nucleation, growth, switching, and phase transformations. Furthermore, because the electro-optic coefficients are directly captured from the evolution of the microstructure, the framework captures both the intrinsic and extrinsic effects arising from domain wall motion and polarization reorientation[39].

The simulations demonstrate that the local electro-optic coefficients are strongly modified by interactions between neighboring domains, which change depending upon the applied electric field direction. From this, we may conclude that the effective electro-optic response of a multidomain film should not be interpreted as a volume average of bulk crystal tensors because the ferroelectric domain interactions and domain walls collectively generate an emergent electro-optic response that can greatly exceed the values predicted for ideal single-domain crystals. This interpretation becomes even more problematic when considering the potential for other ferroelectric phases, such as low-symmetry monoclinic phases, which are not present in the bulk phase diagram[3,40].

Our simulations show that the largest electro-optic response occurs near the orthorhombic–rhombohedral phase boundary, which is bridged by a low-symmetry monoclinic B phase (Figure 6a-c). These results suggest that engineering regions of structural instability and phase competition are key elements to produce a large electro-optic response. Applying electric fields to the system provides another avenue to realize these large electro-optic effects, which is seen in the large transient electro-optic responses as the domain structure reconfigures during the switching process (Figure 4c and Figure 5c). Although these transient states may be unstable, future work may develop techniques to stabilize them, analogous to efforts to stabilize negative capacitance[41].

Beyond electro-optic effects, our work establishes a general platform for simulating light-matter interactions in complex ferroelectric microstructures. This same methodology can be utilized to predict the effective refractive index, optical scattering[42], nonlinear optical interactions, waveguide devices, plasmonics[34], metamaterials[43], and frequency-dependent properties from the quasistatic to the optical frequencies. More broadly, this work provides new pathways towards the computational design of optical functionalities at the mesoscale, which may catalyze the development of improved photonic devices and new device concepts that harness the unique physics of ferroelectric materials.

## Conclusion

In this work, we formulated a phase-field model for the optical properties of ferroelectric materials. The approach extends the phase-field model of ferroelectrics by separating the total polarization into its lattice and electronic contributions, thereby capturing the full temperature- and wavelength-dependent anisotropic optical properties and resolving the temporal dynamics of the local optical properties in complex mesoscale structures. We applied this approach to $BaTiO_3$-based thin films and demonstrated that the model reproduces the experimentally observed temperature and orientation dependence of the electro-optic response. We revealed that the local optical properties are strongly dependent on the ferroelectric microstructures, with local electro-optic effects exceeding 4000 pm/V near domain boundaries, and a transient film-averaged response during switching that increases with field angle to ~6000 pm/V. More broadly, this work establishes a general method for studying light-matter interactions in complex ferroelectric microstructures. Beyond the electro-optic effect, the model can be extended to investigate optical scattering, nonlinear optical phenomena, waveguides, and the frequency-dependent electro-optic response, providing a new pathway for the computational design of ferroelectric materials for photonic devices.

## Acknowledgements

The work was primarily supported as part of the Computational Materials Sciences Program funded by the U.S. Department of Energy, Office of Science, Basic Energy Sciences, under Award DE-SC0020145 (A.R., V.G., and L.Q.C.). A.R. acknowledges the support of the National Science Foundation Graduate Research Fellowship Program under Grant No. DGE1255832. A.F. acknowledges the support of the National Science Foundation through Penn State's Sustainable Physics Research Experiences for Undergraduates under Award PHY-2349159 and its Materials Research Science and Engineering Center under Award DMR-2011839. The phase-field simulations in this work were conducted with the advanced computing resources provided by Texas A&M High Performance Research Computing, through allocation MAT250123 from the Advanced Cyberinfrastructure Coordination Ecosystem: Services & Support (ACCESS) program, which is supported by National Science Foundation grants #2138259, #2138286, #2138307, #2137603, and #2138296.

## Data Availability

The data that supports the findings of this study are available within the article. Datasets used for the generation of the figures are available at DOI: 10.5281/zenodo.21361651.

## Code Availability

The MuPRO code package used in this work to carry out the phase-field simulations can be obtained after a user license is authorized by the MuPRO company (https://muprosoftware.com).

## Competing Interests

Long-Qing Chen has a financial interest in MuPRO, LLC, a company which licenses and markets the software package used in this research.

## Author Contributions

A.R. conceived the study with V.G. and L.Q.C., developed the theoretical framework and phase-field model, performed the simulations, analyzed the results, and wrote the manuscript. A.F. helped establish and troubleshoot the simulation procedure. V.G. and L.Q.C. supervised the research. All authors discussed the results and contributed to revising the manuscript.

## Methods

Based on the thermodynamic theory of optical properties, we may formulate a phase-field model, where the thermodynamics of the ferroelectric system is described by its free energy functional,

$$F = \int\left[f^L\left(T, P_i^L\right) + f^e\left(T, P_i^e\right) + f^{L-e}\left(P_i^L, P_i^e\right) + f^{grad}\left(\nabla_i P_j^L\right) + f^{elas}\left(P_i^L, P_i^e, \varepsilon_{ij}\right) + f^{elec}\left(P_i^L, P_i^e, E_i\right)\right]dV, \quad (16)$$

where $P_i^L(x_i, t)$ is the lattice polarization, $P_i^e(x_i, t)$ is the electronic polarization and $\sigma_{ij}(x, t)$ is the stress[26]. The respective evolution of the lattice polarization and electronic polarization is solved separately. The lattice polarization is computed following the relaxational approximation where we assume relaxational kinetics for the lattice polarization, and the mechanical displacement, electrical potential, and electrical polarization instantaneously reach equilibrium, leading to the time-dependent Ginzburg-Landau equation

$$\gamma_{ij}^L \frac{\partial P_j^L}{\partial t} = -\frac{\delta F}{\delta P_i^L}, \quad (17)$$

where $-\frac{\delta F}{\delta P_i^L}$ is the thermodynamic driving force for the temporal evolution of the lattice polarization. In the simulation, the time step is normalized such that the kinetic coefficient is equal to 1 in reduced units. The normalized timestep is $\Delta t^*$=0.005. We use the local stress and lattice polarization distribution as inputs to the electronic polarization dynamic equation

$$\mu_{ij}^e \frac{\partial^2 P_j^e}{\partial t^2} + \gamma_{ij}^e \frac{\partial P_j^e}{\partial t} = -\frac{\delta F}{\delta P_i^e}, \quad (18)$$

where $-\frac{\delta F}{\delta P_i^e}$ is the thermodynamic driving force for the temporal evolution of the electronic polarization. To compute the local electronic dielectric susceptibility, we solve equation 18 assuming a small periodic electric field is applied, which yields an analytical solution

$$\widetilde{\chi_{ij}^{e,1}}(x_i, \omega) = \left[B_{ij}^e(x_i) - \epsilon_0\left(i\omega\gamma_{ij}^e + \omega^2\mu_{ij}^e\right)\right]^{-1}, \quad (19)$$

which is related to the curvature of the free-energy landscape with respect to the electronic polarization by, $B_{ij}^e(x_i) = \epsilon_0\left(\frac{\partial^2 f}{\partial P_i^e, \partial P_j^e}\right)$. Solving equation 19 (assuming $\lambda = 1550\ nm$) we may compute the local refractive indices from the local lattice polarization and stress distribution, thereby naturally including the electro-optic effect and potential piezo-optic effects. As the lattice polarization distribution evolves in response to an applied electric field, the corresponding change to the electronic dielectric susceptibility and the refractive index is computed.

Here $f^L\left(P_i^L\right)$ describes the intrinsic stability of the lattice polarization compared to the high symmetry phase ($m\bar{3}m$) as a Taylor expansion of the polarization about the high symmetry phase, this is equivalent to the Landau free energy density:

$$f^L(T, P_i^L) = f_o + a_{ij}(T)P_i^L P_j^L + a_{ijkl}P_i^L P_j^L P_k^L P_l^L + a_{ijklmn}P_i^L P_j^L P_k^L P_l^L P_m^L P_n^L + a_{ijklmnop}P_i^L P_j^L P_k^L P_l^L P_m^L P_n^L P_o^L P_p^L, \quad (20)$$

where $a_{ij}$, $a_{ijkl}$, $a_{ijklmn}$ and $a_{ijklmnop}$ are the dielectric stiffness coefficients measured under constant stress conditions. For $BaTiO_3$, an $8^{th}$ order expansion is used to describe the stability of the lattice polarization. The thermodynamic coefficients are adapted from ref. [44] to describe the effects of quantum fluctuations at cryogenic temperatures[45].

The intrinsic free energy density of the electronic polarization, in the absence of the lattice polarization, is described by

$$f^e(T, P_i^e) = \frac{1}{2\epsilon_0} B_{ij}^{e,ref}(T) P_i^e P_j^e, \quad (21)$$

where $B_{ij}^{ref}(T)$ is related to the refractive index of the equivalent high symmetry phase. The coupling energy density between the lattice and electronic polarization which determines the electro-optic effect is given by

$$f^{L-e}(P_i^L, P_i^e) = \frac{1}{2\epsilon_0} g_{ijkl}^{LL} P_l^L P_k^L P_i^e P_j^e, \quad (22)$$

where $g_{ijkl}^{LL}$ relates the lattice polarization to the refractive index.
The gradient energy density is represented by

$$f_{grad} = \frac{1}{2} G_{ijkl} \frac{\partial P_i^L}{\partial x_j} \frac{\partial P_k^L}{\partial x_l}, \quad (23)$$

where $G_{ijkl}$ is the gradient energy tensor where the non-zero coefficients are chosen to be $G_{11} = 0.6, G_{12} = -0.6$ and $G_{44} = 0.6$, and the units are normalized by $\alpha_1 l_o^2$, where $\alpha_1$ is the first Landau expansion coefficient and $l_o$ is chosen as 1 nm per grid point.
The elastic energy is given by

$$f^{elas}(P_i^L, P_i^e, \sigma_{ij}) = \tfrac{1}{2} C_{ijkl}(\varepsilon_{ij} - \varepsilon_{ij}^o)(\varepsilon_{kl} - \varepsilon_{kl}^o), \quad (24)$$

where $C_{ijkl}$ is the elastic stiffness, $\varepsilon_{ij}$ is the total strain, and $\varepsilon_{ij}^o = Q_{ijkl}P_k^L P_l^L + \frac{1}{2\epsilon_0}\pi_{ijkl}P_k^e P_l^e$ is the eigenstrain, where $Q_{ijkl}$ is the electrostrictive coefficient and $\pi_{ijkl}$ is the piezo-optic tensor for the high-symmetry phase. The total strain is solved for a thin film boundary condition assuming that the strain relaxes to its equilibrium value at each time step; for simplicity we ignore the contribution of the electronic polarization to the eigenstrain. Further details on solving the elasticity may be found in ref. [46].
The electrostatic energy is given by

$$f^{elec} = -E_i P_i^L - E_i P_i^e - \tfrac{1}{2}\epsilon_0 \kappa_{ij}^b E_i E_j, \quad (25)$$

where $\epsilon_0$ is the vacuum permittivity and $\kappa_{ij}^b$ is the background dielectric constant. For simplicity to compute evolution of the lattice polarization we only include the linear contribution to electronic polarization and add it to the background dielectric constant yielding $f^{elec} = -E_i P_i^L - \frac{1}{2}\epsilon_0 \kappa_{ij}^b E_i E_j$, where $\kappa_{ij}^b = 10$ and is isotropic. Here $\kappa_{ij}^b$ contains contributions from the electronic contribution and the vacuum and other hard modes. A table containing all coefficients used in the phase-field simulations is given in the supplementary material.

For the simulations, a system size of $256\ \Delta x_1 \times 256\ \Delta x_2\ \times 36\ \Delta x_3$ is used. There are $12\ \Delta x_3$ grid points for the substrate where the elastic constants are assumed to be the same as the ferroelectric film, and there are $4\ \Delta x_3$ grid points acting as an air layer above the film. The thickness of $BaTiO_3$ is set to be 20 $\Delta x_3$. Periodic boundary conditions are used along the in-plane lateral directions, and natural boundary conditions are used along the film-substrate and film-air interfaces. The simulations in Figures 3–5 are performed at 350 K under zero misfit strain, representing a fully relaxed film and providing a strain-free reference state in which the microstructural contributions to the electro-optic response can be isolated. For the temperature-dependent comparison with the measurements of Eltes et al.[18] (Figure 6), we impose a small biaxial tensile strain of 0.15%, representing $BaTiO_3$ films integrated on Si, which relax nearly completely at the growth temperature but retain a small residual tensile strain upon cooling due to the thermal expansion mismatch between $BaTiO_3$ and Si. This strain state stabilizes the experimentally observed in-plane polarization orientation of $BaTiO_3$/Si films. For simplicity, the misfit strain is held constant over the simulated temperature range; the temperature dependence of the residual thermal strain is small compared to the spontaneous strain of the ferroelectric phases and is neglected.

The simulations begin with an initial condition of $P_i^L(x_i, t = 0) = \Delta P^{noise}(x_i, t = 0)$, where $|P^{noise}| = 0.1\ \mathrm{C\ m^{-2}}$, and relax for 20,000 time steps. To simulate the evolution of the polarization under an applied electric field and the corresponding electro-optic effect, a uniform electric field is applied along the $[100]_{pc}$ direction ranging from -50 kV cm$^{-1}$ to 50 kV cm$^{-1}$ which varies in time with a sinewave, with a period of 140,000 time steps. The linear electro-optic effect is found by taking the average of the refractive index over the ferroelectric material as a function of an applied electric field and fitting to a cubic polynomial centred at zero and extracting the linear contribution. The lattice–polarization evolution was computed using the Mu-Pro phase-field simulation package[47].

## References

1. Boyd, R. W. *Nonlinear Optics*. (Academic Press, 2020).

2. Wang, C., Zhang, M., Chen, X., Bertrand, M., Shams-Ansari, A., Chandrasekhar, S., Winzer, P. & Lončar, M. Integrated lithium niobate electro-optic modulators operating at CMOS-compatible voltages. *Nature* **562**, 101–104 (2018).

3. Suceava, A., Hazra, S., Ross, A., Philippi, I. R., Sotir, D., Brower, B., Ding, L., Zhu, Y., Zhang, Z., Sarkar, H., Sarker, S., Yang, Y., Sarker, S., Stoica, V. A., Schlom, D. G., Chen, L.-Q. & Gopalan, V. Colossal Cryogenic Electro-Optic Response Through Metastability in Strained BaTiO3 Thin Films. *Advanced Materials* **38**, e07564 (2026).

4. Ross, A., Hazra, S., Suceava, A., Sotir, D., Schlom, D. G., Gopalan, V. & Chen, L.-Q. Quantum Saturation of the Electro-Optic Effect. Preprint at https://doi.org/10.48550/arXiv.2603.23486 (2026).

5. Warner, H. K., Holzgrafe, J., Yankelevich, B., Barton, D., Poletto, S., Xin, C. J., Sinclair, N., Zhu, D., Sete, E., Langley, B., Batson, E., Colangelo, M., Shams-Ansari, A., Joe, G., Berggren, K. K., Jiang, L., Reagor, M. J. & Lončar, M. Coherent control of a superconducting qubit using light. *Nat. Phys.* 1–8 (2025) doi:10.1038/s41567-025-02812-0.

6. Anderson, C. P., Scuri, G., Chan, A., Eun, S., White, A. D., Ahn, G. H., Jilly, C., Safavi-Naeini, A., Van Gasse, K., Li, L. & Vučković, J. Quantum critical electro-optic and piezo-electric nonlinearities. *Science* **390**, 394–399 (2025).

7. Lejman, M., Vaudel, G., Infante, I. C., Chaban, I., Pezeril, T., Edely, M., Nataf, G. F., Guennou, M., Kreisel, J., Gusev, V. E., Dkhil, B. & Ruello, P. Ultrafast acousto-optic mode conversion in optically birefringent ferroelectrics. *Nat Commun* **7**, 12345 (2016).

8. Yudistira, D., Janner, D., Benchabane, S. & Pruneri, V. Integrated acousto-optic polarization converter in a ZX-cut $LiNbO_3$ waveguide superlattice. *Opt. Lett., OL* **34**, 3205–3207 (2009).

9. Boes, A., Chang, L., Langrock, C., Yu, M., Zhang, M., Lin, Q., Lončar, M., Fejer, M., Bowers, J. & Mitchell, A. Lithium niobate photonics: Unlocking the electromagnetic spectrum. *Science* **379**, eabj4396 (2023).

10. Geler-Kremer, J., Eltes, F., Stark, P., Stark, D., Caimi, D., Siegwart, H., Jan Offrein, B., Fompeyrine, J. & Abel, S. A ferroelectric multilevel non-volatile photonic phase shifter. *Nat. Photon.* **16**, 491–497 (2022).

11. Onodera, T., Stein, M. M., Ash, B. A., Sohoni, M. M., Bosch, M., Yanagimoto, R., Jankowski, M., McKenna, T. P., Wang, T., Shvets, G., Shcherbakov, M. R., Wright, L. G. & McMahon, P. L. Arbitrary control over multimode wave propagation for machine learning. *Nat. Phys.* **22**, 164–171 (2026).

12. Myers, L. E., Eckardt, R. C., Fejer, M. M., Byer, R. L., Bosenberg, W. R. & Pierce, J. W. Quasi-phase-matched optical parametric oscillators in bulk periodically poled $LiNbO_3$. *J. Opt. Soc. Am. B, JOSAB* **12**, 2102–2116 (1995).

13. Yanagimoto, R., Ash, B. A., Sohoni, M. M., Stein, M. M., Zhao, Y., Presutti, F., Jankowski, M., Wright, L. G., Onodera, T. & McMahon, P. L. Programmable on-chip nonlinear photonics. *Nature* **649**, 330–337 (2026).

14. Denev, S. A., Lummen, T. T. A., Barnes, E., Kumar, A. & Gopalan, V. Probing Ferroelectrics Using Optical Second Harmonic Generation. *Journal of the American Ceramic Society* **94**, 2699–2727 (2011).

15. Chelladurai, D., Kohli, M., Winiger, J., Moor, D., Messner, A., Fedoryshyn, Y., Eleraky, M., Liu, Y., Wang, H. & Leuthold, J. Barium titanate and lithium niobate permittivity and Pockels coefficients from megahertz to sub-terahertz frequencies. *Nat. Mater.* **24**, 868–875 (2025).

16. Zgonik, M., Bernasconi, P., Duelli, M., Schlesser, R., Günter, P., Garrett, M. H., Rytz, D., Zhu, Y. & Wu, X. Dielectric, elastic, piezoelectric, electro-optic, and elasto-optic tensors of BaTiO3 crystals. *Phys. Rev. B* **50**, 5941–5949 (1994).

17. Abel, S., Stöferle, T., Marchiori, C., Rossel, C., Rossell, M. D., Erni, R., Caimi, D., Sousa, M., Chelnokov, A., Offrein, B. J. & Fompeyrine, J. A strong electro-optically active lead-free ferroelectric integrated on silicon. *Nat Commun* **4**, 1671 (2013).

18. Eltes, F., Villarreal-Garcia, G. E., Caimi, D., Siegwart, H., Gentile, A. A., Hart, A., Stark, P., Marshall, G. D., Thompson, M. G., Barreto, J., Fompeyrine, J. & Abel, S. An integrated optical modulator operating at cryogenic temperatures. *Nat. Mater.* **19**, 1164–1168 (2020).
19. Veithen, M., Gonze, X. & Ghosez, Ph. Nonlinear optical susceptibilities, Raman efficiencies, and electro-optic tensors from first-principles density functional perturbation theory. *Phys. Rev. B* **71**, 125107 (2005).
20. Fredrickson, K. D., Vogler-Neuling, V. V., Kormondy, K. J., Caimi, D., Eltes, F., Sousa, M., Fompeyrine, J., Abel, S. & Demkov, A. A. Strain enhancement of the electro-optical response in BaTi O 3 films integrated on Si ( 001 ). *Phys. Rev. B* **98**, 075136 (2018).
21. Kim, I., Paoletta, T. & Demkov, A. A. Nature of electro-optic response in tetragonal ${\mathrm{BaTiO}}_{3}$. *Phys. Rev. B* **108**, 115201 (2023).
22. Paillard, C., Prokhorenko, S. & Bellaiche, L. Strain engineering of electro-optic constants in ferroelectric materials. *npj Comput Mater* **5**, 6 (2019).
23. DiDomenico, M., Jr. & Wemple, S. H. Oxygen-Octahedra Ferroelectrics. I. Theory of Electro-optical and Nonlinear optical Effects. *Journal of Applied Physics* **40**, 720–734 (1969).
24. Wemple, S. H. & DiDomenico, M. Theory of the Elasto-Optic Effect in Nonmetallic Crystals. *Phys. Rev. B* **1**, 193–202 (1970).
25. Chen, L.-Q. Phase-Field Method of Phase Transitions/Domain Structures in Ferroelectric Thin Films: A Review. *Journal of the American Ceramic Society* **91**, 1835–1844 (2008).
26. Ross, A., Ali, M. S. M. M., Saha, A., Zu, R., Gopalan, V., Dabo, I. & Chen, L.-Q. Thermodynamic theory of linear optical and electro-optical properties of ferroelectrics. *Phys. Rev. B* **111**, 085109 (2025).
27. Li, W., Landis, C. M. & Demkov, A. A. Domain morphology and electro-optic effect in Si-integrated epitaxial $\mathrm{BaTi}{\mathrm{O}}_{3}$ films. *Phys. Rev. Mater.* **6**, 095203 (2022).

28. Wong, C., Teng, Y. Y., Ashok, J. & Varaprasad, P. L. H. Barium Titanate (BaTiO3). in *Handbook of Optical Constants of Solids* (ed. Palik, E. D.) 789–803 (Academic Press, Burlington, 1997). doi:10.1016/B978-012544415-6.50080-7.

29. Cochran, W. Crystal stability and the theory of ferroelectricity. *Advances in Physics* **9**, 387–423 (1960).

30. Pertsev, N. A. & Arlt, G. Forced translational vibrations of 90° domain walls and the dielectric dispersion in ferroelectric ceramics. *J. Appl. Phys.* **74**, 4105–4112 (1993).

31. Devonshire, A. F. Theory of ferroelectrics. *Advances in Physics* **3**, 85–130 (1954).

32. Ginzburg, V. L. Polarizasyon and piezoelectric effect in BaTiO3 near the ferroelectric transition point. *Zh. Eksp. Teor. Fiz* **19**, 36–41 (1949).

33. Yang, T., Wang, B., Hu, J.-M. & Chen, L.-Q. Domain Dynamics under Ultrafast Electric-Field Pulses. *Phys. Rev. Lett.* **124**, 107601 (2020).

34. Abel, S., Eltes, F., Ortmann, J. E., Messner, A., Castera, P., Wagner, T., Urbonas, D., Rosa, A., Gutierrez, A. M., Tulli, D., Ma, P., Baeuerle, B., Josten, A., Heni, W., Caimi, D., Czornomaz, L., Demkov, A. A., Leuthold, J., Sanchis, P. & Fompeyrine, J. Large Pockels effect in micro- and nanostructured barium titanate integrated on silicon. *Nature Mater* **18**, 42–47 (2019).

35. Demkov, A. A., Bajaj, C., Ekerdt, J. G., Palmstrøm, C. J. & Ben Yoo, S. J. Materials for emergent silicon-integrated optical computing. *Journal of Applied Physics* **130**, 070907 (2021).

36. Alexander, K., Benyamini, A., Black, D., Bonneau, D., Burgos, S., Burridge, B., Cable, H., Campbell, G., Catalano, G., Ceballos, A., Chang, C.-M., Choudhury, S. S., Chung, C., Danesh, F., Dauer, T., Davis, M., Dudley, E., Er-Xuan, P., Fargas, J., Farsi, A., Fenrich, C., Frazer, J., Fukami, M., Ganesan, Y., Gibson, G., Gimeno-Segovia, M., Goeldi, S., Goley, P., Haislmaier, R., Halimi, S., Hansen, P., Hardy, S., Horng, J., House, M., Hu, H., Jadidi, M., Jain, V., Johansson, H., Jones, T., Kamineni, V., Kelez, N., Koustuban, R., Kovall, G., Krogen, P., Kumar, N., Liang, Y., LiCausi, N., Llewellyn, D., Lokovic, K., Lovelady, M., Manfrinato, V. R., Melnichuk, A., Mendoza, G., Moores, B., Mukherjee, S., Munns, J., Musalem, F.-X., Najafi, F., O'Brien, J. L., Ortmann, J. E., Pai, S., Park,

B., Peng, H.-T., Penthorn, N., Peterson, B., Peterson, G., Poush, M., Pryde, G. J., Ramprasad, T., Ray, G., Rodriguez, A. V., Roxworthy, B., Rudolph, T., Saunders, D. J., Shadbolt, P., Shah, D., Bahgat Shehata, A., Shin, H., Sinsky, J., Smith, J., Sohn, B., Sohn, Y.-I., Son, G., Souza, M. C. M. M., Sparrow, C., Staffaroni, M., Stavrakas, C., Sukumaran, V., Tamborini, D., Thompson, M. G., Tran, K., Triplett, M., Tung, M., Veitia, A., Vert, A., Vidrighin, M. D., Vorobeichik, I., Weigel, P., Wingert, M., Wooding, J., Zhou, X., & PsiQuantum Team. A manufacturable platform for photonic quantum computing. *Nature* 1–3 (2025) doi:10.1038/s41586-025-08820-7.

37. Castera, P., Tulli, D., Gutierrez, A. M. & Sanchis, P. Influence of $BaTiO_3$ ferroelectric orientation for electro-optic modulation on silicon. *Opt. Express, OE* **23**, 15332–15342 (2015).
38. Vasudevan, A. T. & Selvaraja, S. K. Domain effects on the electro-optic properties of thin-film barium titanate. *Opt. Mater. Express, OME* **13**, 956–965 (2023).
39. Damjanovic, D. Ferroelectric, dielectric and piezoelectric properties of ferroelectric thin films and ceramics. *Rep. Prog. Phys.* **61**, 1267 (1998).
40. Lummen, T. T. A., Gu, Y., Wang, J., Lei, S., Xue, F., Kumar, A., Barnes, A. T., Barnes, E., Denev, S., Belianinov, A., Holt, M., Morozovska, A. N., Kalinin, S. V., Chen, L.-Q. & Gopalan, V. Thermotropic phase boundaries in classic ferroelectrics. *Nat Commun* **5**, 3172 (2014).
41. Yadav, A. K., Nguyen, K. X., Hong, Z., García-Fernández, P., Aguado-Puente, P., Nelson, C. T., Das, S., Prasad, B., Kwon, D., Cheema, S., Khan, A. I., Hu, C., Íñiguez, J., Junquera, J., Chen, L.-Q., Muller, D. A., Ramesh, R. & Salahuddin, S. Spatially resolved steady-state negative capacitance. *Nature* **565**, 468–471 (2019).
42. Townsend, J., Picinini, E. C. & Sousa, R. de. Scattering-Induced Loss in Ferroelectric Photonic Devices. Preprint at https://doi.org/10.48550/arXiv.2605.04353 (2026).
43. Karvounis, A., Timpu, F., Vogler-Neuling, V. V., Savo, R. & Grange, R. Barium Titanate Nanostructures and Thin Films for Photonics. *Advanced Optical Materials* **8**, 2001249 (2020).
44. Li, Y. L., Cross, L. E. & Chen, L. Q. A phenomenological thermodynamic potential for BaTiO3 single crystals. *Journal of Applied Physics* **98**, 064101 (2005).

45. Barrett, J. H. Dielectric Constant in Perovskite Type Crystals. *Phys. Rev.* **86**, 118–120 (1952).
46. Li, Y. L., Hu, S. Y., Liu, Z. K. & Chen, L. Q. Effect of substrate constraint on the stability and evolution of ferroelectric domain structures in thin films. *Acta Materialia* **50**, 395–411 (2002).
47. Cheng, X. MUPRO. *MUPRO* https://mupro.co/.

## Supplementary Information:

**Supplementary Note 1: Parameters in the thermodynamic free-energy function for $BaTiO_3$**

For $BaTiO_3$, we use the cubic phase ($m\bar{3}m$) as our high symmetry reference state and employ an 8$^{th}$-order Landau expansion to describe the relative stability of the lattice polarization compared to the cubic reference state. The coefficients used for this paper are adjusted from Li *et al.*[44] and Ross *et al.*[26] and to include the effect of cryogenic fluctuations and are given in Table S1.

**Table S1.** Coefficients in the thermodynamic free energy function and equation of motion for $BaTiO_3$

| | | | |
|---|---|---|---|
| $g^{LL}_{1111}$ | $18.5 \times 10^{-2} (\mathrm{m^4/C^2})$ | $p_{1111}$ | 0.5328 (Unitless) |
| $g^{LL}_{1122}$ | $2.5 \times 10^{-2} (\mathrm{m^4/C^2})$ | $p_{1122}$ | 0.1584 (Unitless) |
| $g^{LL}_{1212}$ | $12.85 \times 10^{-2} (\mathrm{m^4/C^2})$ | $p_{1212}$ | −0.432 (Unitless) |
| $\mu_e$ | $35.5 \times 10^{-23} \left(\frac{\mathrm{Kg}}{\mathrm{m}}\frac{\mathrm{m^4}}{\mathrm{C^2}}\right)$ | $\gamma_e$ | $3 \times 10^{-9} \left(\frac{\mathrm{Kg}}{\mathrm{ms}}\frac{\mathrm{m^4}}{\mathrm{C^2}}\right)$ |
| $a_{11}$ | $a_0\, T_s \left(\coth\left(\frac{T_s}{T}\right) - \coth(\frac{T_s}{T_c})\right)$ | $B^{e,ref}_{ij}(T_0)$ | 0.2356 (Unitless) |
| $T_c$ | 388 K | $T_0$ | 398 K |
| $T_s$ | 54 K | $a_0$ | $4.124 \times 10^{5} \left(\frac{\mathrm{J}}{\mathrm{m^3}}\ \frac{\mathrm{m^4}}{\mathrm{K\,C^2}}\right)$ |
| $a_{1111}$ | $-2.097 \times 10^{8} \left(\frac{\mathrm{J}}{\mathrm{m^3}}\ \frac{\mathrm{m^8}}{\mathrm{C^4}}\right)$ | $\alpha_{11}$ | $2.657 \times 10^{-5} \left(\frac{1}{\mathrm{K}}\right)$ |
| $a_{1122}$ | $7.974 \times 10^{8} \left(\frac{\mathrm{J}}{\mathrm{m^3}}\ \frac{\mathrm{m^8}}{\mathrm{C^4}}\right)$ | $a_{11111111}$ | $3.863 \times 10^{10} \left(\frac{\mathrm{J}}{\mathrm{m^3}}\ \frac{\mathrm{m^{16}}}{\mathrm{C^8}}\right)$ |
| $a_{111111}$ | $1.294 \times 10^{9} \left(\frac{\mathrm{J}}{\mathrm{m^3}}\ \frac{\mathrm{m^{12}}}{\mathrm{C^6}}\right)$ | $a_{11111122}$ | $2.529 \times 10^{10} \left(\frac{\mathrm{J}}{\mathrm{m^3}}\ \frac{\mathrm{m^{16}}}{\mathrm{C^8}}\right)$ |
| $a_{111122}$ | $-1.95 \times 10^{9} \left(\frac{\mathrm{J}}{\mathrm{m^3}}\ \frac{\mathrm{m^{12}}}{\mathrm{C^6}}\right)$ | $a_{11112222}$ | $1.637 \times 10^{10} \left(\frac{\mathrm{J}}{\mathrm{m^3}}\ \frac{\mathrm{m^{16}}}{\mathrm{C^8}}\right)$ |
| $a_{112233}$ | $-2.509 \times 10^{9} \left(\frac{\mathrm{J}}{\mathrm{m^3}}\ \frac{\mathrm{m^{12}}}{\mathrm{C^6}}\right)$ | $a_{11112233}$ | $1.367 \times 10^{10} \left(\frac{\mathrm{J}}{\mathrm{m^3}}\ \frac{\mathrm{m^{16}}}{\mathrm{C^8}}\right)$ |
| $C_{11}$ | $1.78 \times 10^{11}$ Pa | $C_{44}$ | $1.22 \times 10^{11}$ Pa |
| $C_{12}$ | $0.964 \times 10^{11}$ Pa | $Q_{11}$ | $0.11\ \frac{\mathrm{m^4}}{\mathrm{C^2}}$ |
| $Q_{12}$ | $-0.040\ \frac{\mathrm{m^4}}{\mathrm{C^2}}$ | $Q_{44}$ | $0.065\ \frac{\mathrm{m^4}}{\mathrm{C^2}}$ |

**Supplementary Note 2: Influence of Poling on Ferroelectric Domain Structures**

Electrical poling modifies the stability of different ferroelectric phases by favoring domains with polarization vectors aligned with the electric field. In general, this process will be affected by both the ferroelectric phase and the potential mesoscale structures that they organize into. In ferroelectric superdomains, individual domains may not fully align with the electric field, but the averaged polarization vector across the superdomain may instead align with the electric field. After the electric field is removed, the material often stays in this aligned state, thereby retaining a memory of the poling process, which influences the temperature-dependent properties.

As shown in Figure S1, the poled states generally exhibit a lower fraction of monoclinic phases and a higher fraction of rhombohedral, orthorhombic, and tetragonal phases. Poling also modifies the apparent phase transition temperature. In particular, the tetragonal to orthorhombic phase transition shifts to lower temperatures, with the magnitude of the temperature shift increasing as the poling direction shifts from 0° to 45°. A similar, but less pronounced shift is also present near the orthorhombic to rhombohedral phase boundary.

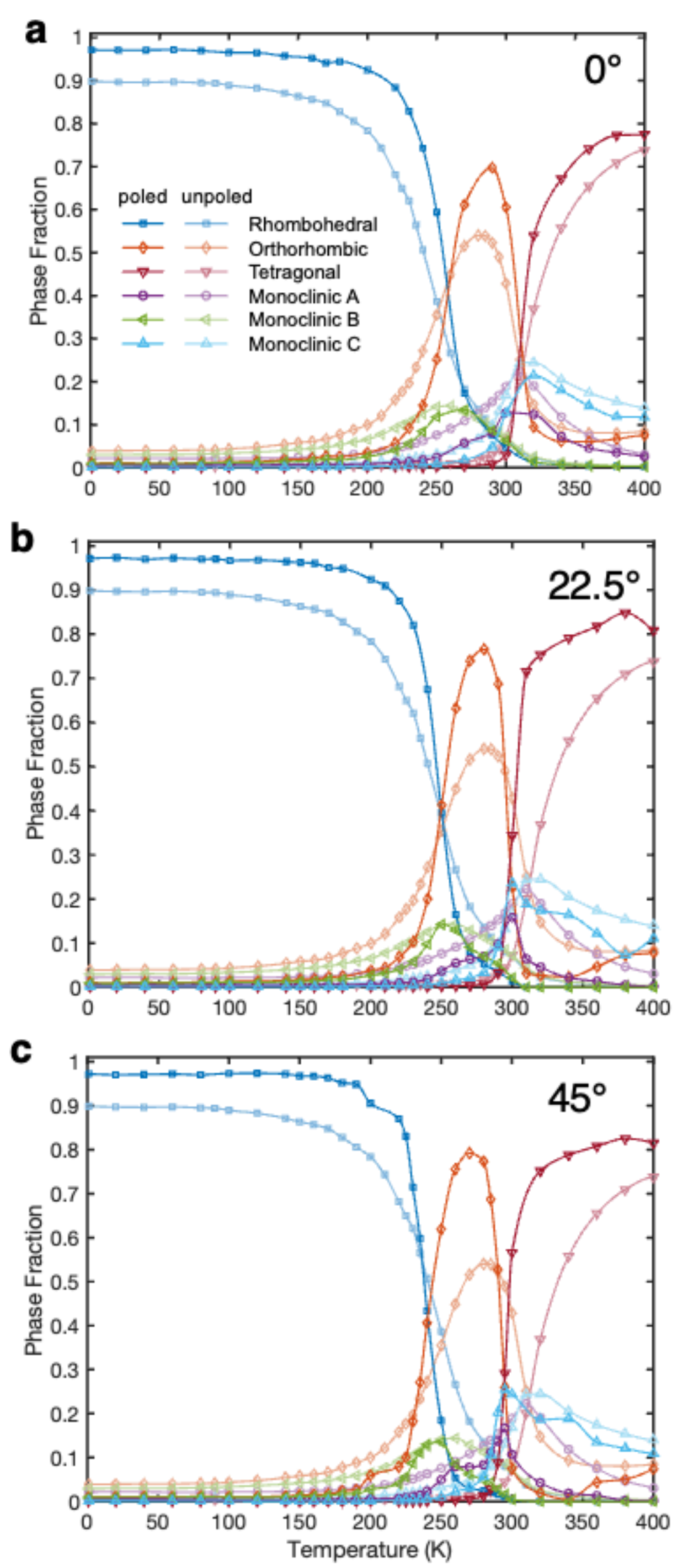

**Supplementary Figure 1.** Influence of the electrical poling direction on the phase fractions. The electric field is applied along the 0°, 22.5°, and 45° directions. Lighter curves correspond to the unpoled state before the application of the electric field, while darker curves correspond to the remanent state after applying a negative bias and returning to zero applied electric field.